\documentclass[aps,twocolumn,nofootinbib]{revtex4-1}
\usepackage{amssymb}
\usepackage{amsmath}
\usepackage{amstext}
\usepackage{amsfonts}
\usepackage{bbold}
\usepackage{slashed}

\def\l{\left}
\def\r{\right}
\def\fr{\frac}
\def\la{\label}
\def\d{\partial}

\def\be{\begin{eqnarray}}
\def\ee{\end{eqnarray}}
\newcommand{\p}{\bar{P}}  
\newcommand{\J}{\bar{J}}  
\newcommand{\D}{{\mathcal D}}

\begin{document}

\title{Relativistic Fluids, Superfluids, Solids and Supersolids from a Coset Construction} 

\author{Alberto Nicolis}
\affiliation{Physics Department and Institute for Strings, Cosmology, and Astroparticle Physics,\\
  Columbia University, New York, NY 10027, USA}
\author{Riccardo Penco}
\affiliation{Physics Department and Institute for Strings, Cosmology, and Astroparticle Physics,\\
  Columbia University, New York, NY 10027, USA}
\author{Rachel A. Rosen}
\affiliation{Physics Department and Institute for Strings, Cosmology, and Astroparticle Physics,\\
  Columbia University, New York, NY 10027, USA}

\begin{abstract}
We provide a systematic coset construction of the  effective field theories governing the low-energy dynamics of relativistic fluids and solids, and of their `super' counterparts. These effective theories agree with those previously derived via different techniques. 
As an application of our methods, we re-derive
the  Wess-Zumino term relevant for anomalous charge-carrying fluids in (1+1) dimensions.
\end{abstract}

\maketitle

\section{Introduction}


Hydrodynamics is usually studied at the level of its equations of motion, which are nothing but the local conservation laws for energy, momentum, and possibly additional conserved charges carried by the fluid in question. Recently, however, it has been realized that an effective field theory treatment in terms of a local action functional might be more convenient for certain applications. This is partially due to the fact that long-wavelength hydrodynamical modes can be thought of as the Goldstone excitations associated with certain spontaneously broken spacetime symmetries, and effective field theory is at present the most efficient tool we have  to characterize systematically the low-energy dynamics of Goldstone excitations. In fact, this logic was probably first explored  for the dynamics of phonons in solids \cite{Leutwyler:1996er}, and only later generalized to zero-temperature superfluids \cite{Son:2002zn}, supersolids \cite{Son:2005ak}, ordinary fluids \cite{Dubovsky:2005xd,Dubovsky:2011sj},  finite-temperature superfluids \cite{Nicolis:2011cs}, and  supersymmetric fluid systems \cite{Hoyos:2012dh,Andrianopoli:2013dya}.

The power of this new approach lies partially in its economy. One functional of the fields, i.e., the action, encodes all the information about the theory: the equations of motion, the stress-energy tensor, the other conserved currents, quantum phenomena, etc.
The power of this approach also lies in how systematic it is.  The action one writes down  should be the most general local functional compatible
with the symmetries, organized at low-energies as a perturbative expansion in the fields' derivatives.

For Goldstones associated with standard spontaneous symmetry breaking in particle physics, the celebrated coset construction  \cite{Callan:1969sn,Coleman:1969sm} is  the most exhaustive technique we have  to write down such low-energy effective actions. Such a technique has been generalized to spontaneously broken {\em spacetime} symmetries in \cite{Volkov:1973vd,Ogievetsky:1974ab}, but so far it has not been applied directly to the systems of our interests: fluids, solids, and variations thereof.

Our perhaps modest goal in this paper is to carry out this application. The motivation is twofold. On the one hand, we  want to confirm
that the effective theories  that have been written down so far for these systems are indeed the most general ones compatible with the appropriate symmetries to lowest order in the derivative expansion (it turns out that they are). On the other hand, we hope that the formalism we  develop here will turn out to be useful in overcoming certain stumbling blocks that have been encountered in trying to extend these effective theories to higher orders---regarding for instance the inclusion of Wess-Zumino terms in dimensions higher than $(1+1)$ \cite{Dubovsky:2011sk}, and the inclusion of dissipative effects to non-linear order in the hydrodynamical modes \cite{Endlich:2012vt}.

We will start with the coset construction for an ordinary fluid that carries a conserved charge. In a sense that will be clear in what follows, this is the most complicated system among those in our title. Then, by gradually removing symmetries, we will be able to generalize such a construction to the other systems as well. For simplicity we will not consider the supersymmetric versions of our systems. Moreover, when covering solids and supersolids, we will only consider isotropic systems that realize the full $SO(3)$ symmetry rather than one of its discrete subgroups, as would be more appropriate for actual crystals.

To describe a generic system featuring spontaneous symmetry breaking, the only input needed by the coset construction is the symmetry breaking pattern. For the systems that we are interested in, the corresponding symmetry breaking patterns have been discussed extensively in the literature cited above. We will thus not re-derive them here, but rather build on these previous results. It is worth mentioning that all these symmetry breaking patterns feature a crucial interplay between spacetime symmetries and internal ones, whereby {\em all} the unbroken symmetries are linear combinations of both types. This further complicates the already subtle coset construction for broken spacetime symmetries.



\section{Symmetries of Perfect Fluids}
In $D=d+1$ space-time dimensions, the low-energy behavior of perfect fluids can be described using $d$ scalar fields $\phi^I(\vec{x},t)$~\cite{Dubovsky:2011sk,Dubovsky:2011sj}. These fields give the comoving (Lagrangian) coordinates $\phi^I$  of the fluid as a function of the physical spatial coordinates $\vec{x}$ and of time $t$. The action for these scalars is invariant under \emph{internal} volume-preserving diffeomorphisms, i.e., 
\be
\label{vdiff}
\phi^I \rightarrow \xi^I(\phi^J)\,,~~~~\det(\partial\xi^I/\partial\phi^J) =1 \, .
\ee
Physically, such a large internal symmetry group encodes the insensitivity of a perfect fluid's dynamics  to adiabatically slow deformations that do not change the  volume of the individual fluid elements. Among an infinite number of symmetries, this group includes most notably shifts and rotations of the comoving coordinates $\phi^I$.

To describe a perfect fluid with a conserved charge, we introduce an additional scalar field $\phi^0$ that shifts under the associated $U(1)$ symmetry. We demand that the charge  be separately conserved within each volume element of the fluid, which is equivalent to demanding that the charge ``flow with the fluid", $j^\mu = n u^\mu$.  This corresponds to requiring that the action be invariant under shifts of $\phi^0$ that depend on the comoving coordinates $\phi^I$, i.e.
\be
\label{chemshift}
\phi^0 \rightarrow \phi^0+f(\phi^I)\, , 
\ee
where $f$ is an arbitrary function \cite{Dubovsky:2011sj, sibiryakov}. This symmetry is referred to as the ``chemical shift" symmetry.  We call this field $\phi^0$ for notational convenience, but {\em no} internal Lorentz invariance rotating $\phi^0$ into the $\phi^I$'s is implied.  

When the fluid is in equilibrium, its comoving coordinates can be chosen to coincide with the physical coordinates, and $\phi^0$ with time, up to proportionality constants which we omit for simplicity:
\be \la{vacuum}
\langle \phi^i \rangle = x^i\, , ~~~~ \langle \phi^0 \rangle = t \, .
\ee
Such a field configuration spontaneously breaks all of the above internal symmetries as well as boosts, spacetime translations, and spatial rotations, and is left invariant only by a linear combination of internal and space-time translations and rotations. At sufficiently low energies, the only relevant excitations around the equilibrium configuration (\ref{vacuum}) are the Goldstone bosons associated with this symmetry breaking.

In what follows, we will derive a low-energy effective action for these Goldstone bosons. In order to successfully carry out such a construction, we will use an approximation. The internal volume-preserving diffeomorphisms \eqref{vdiff} and the chemical shift \eqref{chemshift} are potentially unwieldy as they are described by an infinite number of generators. To handle this, we restrict the transformation \eqref{vdiff} to constant shifts, generated by $Q_I$, and special linear transformations (i.e., unit determinant $3\times3$ matrices), generated by $M_{IJ}$.  The antisymmetric part of $M_{IJ}$ generates internal rotations and we will denote it by $L_{IJ}$.  Similarly, we will restrict the chemical shift symmetry \eqref{chemshift} to a constant shift, generated by $Q_0$, as well as a shift linear in $\phi^I$, generated by $F_I$.  As we will see, by demanding invariance under this restricted set of symmetries, our action will be accidentally invariant under the full transformations (\ref{vdiff}) and (\ref{chemshift}) to lowest order in derivatives. In fact, at present it is not clear  yet whether the infinite-dimensional symmetries postulated above should survive beyond lowest order in the derivative expansion. For instance,   dissipative effects associated with shear viscosity and heat conduction apparently violate \eqref{vdiff} and \eqref{chemshift} \cite{Endlich:2012vt}. This further indicates that these symmetries might not be fundamental, but only accidental, in which case they should not enter the coset construction.

Finally, we will denote the generators of space-time translations, rotations, and Lorentz boosts by $P_\mu$, $J_{ij}$ and $K_i$ respectively. Therefore, the pattern of symmetry breaking that we will consider can be summarized as follows:
\be
\label{pattern}
\begin{array}{lcl}
\mbox{unbroken} &=&  \left\{
\begin{array}{ll}
\p_\mu \equiv P_\mu+Q_\mu &  \quad\,\,\,\, \mbox{translations} \\
\J_{ij} \equiv J_{ij}+L_{ij}&  \quad\,\,\,\, \mbox{rotations}
\end{array}
\right. 
\\ && \\
\mbox{broken} &=&  \left\{
\begin{array}{ll}
K_i &  \qquad\qquad\quad\quad\,\,   \mbox{boosts} \\
Q_\mu &  \qquad\qquad\quad\quad \,\,  \mbox{constant shifts} \\
F_i &  \qquad\qquad\quad\quad \,\,  \mbox{chemical shifts} \\
M_{ij} &  \qquad\qquad\quad\quad \,\,  \mbox{special linear}
\end{array}
\right. \nonumber
\end{array}
\ee
In what follows,  we will denote the full symmetry group by $G$ and the unbroken subgroup by $H$.  
In $D=4$ dimensions there are 25 generators in total, 18 of which are broken by the field configuration~(\ref{vacuum}).

Notice that since Lorentz invariance is spontaneously broken, the $\mu=0$ and $\mu=i$ components should be treated as independent here and in what follows. Moreover, we have stopped differentiating between the internal $I,J, \dots$ indices and the spatial $i,j, \dots$ ones, since the only index contractions that make sense
at the level of the coset construction are those associated with the unbroken symmetries. So, from here on out all quantities carrying $i,j, \dots$ indices transform as tensors under the unbroken rotations generated by $\bar J_{ij}$.

\section{Coset Construction for Fluids}
\label{fluids}
As per the usual construction~\cite{Callan:1969sn,Coleman:1969sm,Volkov:1973vd,Ogievetsky:1974ab}, we parameterize the space of (left) cosets 
$G/H$ 
by introducing one Goldstone field for each broken generator:
\be
\label{Om}
\Omega(x) = e^{i x^\mu \p_\mu} e^{i \eta^i (x) K_i} e^{i \pi^\mu(x) Q_\mu} e^{i \theta^i(x) F_i}  e^{i \alpha^{ij}(x) M_{ij}} \, .
\ee
In order to construct an effective action that is invariant under the full symmetry group $G$, one considers the Maurer-Cartan form expanded in the basis of generators
\be \la{MC}
\begin{array}{lcl}
\Omega^{-1} d\Omega &\!\!=\!\!& i \omega_P^\mu \p_\mu+i \omega_J^{ij} \J_{ij}+i \omega_K^i K_i \\
&&+i \omega_Q^\mu Q_\mu+i \omega_F^i F_i +i \omega_M^{ij} M_{ij} \, .
\end{array}
\ee
The one-forms $\omega_P^\mu$ are related to the spacetime vielbeins, 
\be
\omega_P^\mu = e_\alpha{}^{\mu} \, dx^\alpha \, .
\ee
The one-forms associated with the broken generators, $\omega_K^i$, $\omega_Q^\mu$, $\omega_F^i$ and $\omega_M^{ij}$, are related to the covariant derivatives of the Goldstone fields,
\be 
\begin{array}{lcl}
\omega_Q^\mu &=& e_\alpha{}^{\nu} \,  \D_\nu \pi^\mu  \, dx^\alpha \, , \\
\omega_F^i &=& e_\alpha{}^{\nu} \,  \D_\nu \theta^i  \, dx^\alpha \, , \\
\omega_M^{ij} &=& e_\alpha{}^{\nu} \,  \D_\nu \alpha^{ij}  \, dx^\alpha \, , \\
\omega_K^i &=& e_\alpha{}^{\nu} \,  \D_\nu \eta^i  \, dx^\alpha \, .
\end{array}
\ee
These forms transform covariantly and can thus be used as building blocks of the invariant Lagrangian.  An action that is constructed to be manifestly invariant under the unbroken group $H$, will automatically be invariant under the full group $G$.

Given the coset parametrization \eqref{Om}, we can use the Poincar\'{e} algebra~\cite{Weinberg:1995mt} as well as the commutation relations
\begin{subequations}
\be
[M_{ij},M_{kl}] &=& i (\delta_{jk} M_{il} - \delta_{il} M_{kj}) ,\\
\l[M_{ij},Q_k\r] &=& i (\delta_{ik} Q_j - \tfrac{1}{d} \delta_{ij} Q_k ) ,\\
\l[M_{ij},F_k\r] &=& i ( \tfrac{1}{d} \delta_{ij} F_k - \delta_{jk} F_i) ,\\
\l[F_i,Q_j\r] &=& i \delta_{ij} Q_0 ,
\ee
\end{subequations}
to calculate the Maurer-Cartan form (\ref{MC}). We obtain
\be
\begin{array}{ccl}
\Omega^{-1}\partial_\mu\Omega\!\!&=&\!\!i \Lambda_\mu{}^{\nu}\p_\nu +e^{-i\eta^iK_i}\partial_\mu e^{i\eta^jK_j}\\
&&\!\!+i[\delta_\mu{}^{0}+\partial_\mu\pi^0+(\delta_\mu{}^{i}+\partial_\mu\pi^i)\theta_i-\Lambda_\mu{}^{0}]Q_0\\
&&\!\!+i[(\delta_\mu{}^{i}+\partial_\mu\pi^i)\xi_i{}^{j}-\Lambda_\mu{}^{j}]Q_j\\
&&\!\!+i\partial_\mu \theta^i \xi_i{}^{j} F_j+e^{-i\alpha^{ij}M_{ij}}\partial_\mu e^{i\alpha^{kl}M_{kl}} \, ,
\end{array} \nonumber
\ee
where $\xi_i{}^{j} = \xi_i{}^{j}(\alpha)$ is a special linear transformation, and $ \Lambda_\mu{}^{\nu} =  \Lambda_\mu{}^{\nu} (\eta)$ is a Lorentz transformation of rapidity $\vec \eta$. 
We have not expanded  the terms involving the  $K$ and $M$ generators.  Since the nested $K$ commutators only generate $K$'s and $J$'s, and the nested $M$ commutators  only generate $M$'s, these will not contribute to the coefficients of the generators $Q_\mu$, which are our primary interest at the moment for reasons that we will explain towards the end of this section. These coefficients give the following covariant derivatives for the $\pi^\mu$ Goldstones:
\begin{subequations}\label{Ds}
\be
\D_\mu \pi^0 &=& -\delta_\mu{}^{0}+\Lambda^\nu{}_{\mu}\left(\partial_\nu \phi^0+\partial_\nu \phi^i \,\theta_i \right) \, , \la{dmupi0}\\
\D_\mu \pi^i &=& -\delta_\mu{}^{i}+\Lambda^\nu{}_{\mu}\partial_\nu\phi^j\,\xi_j{}^{i} \, ,
\ee
\end{subequations}
where we have simplified the notation by introducing the fields $\phi^\mu = x^\mu+\pi^\mu$. It will turn out that these fields are exactly the $\phi^i, \phi^0$ fields  described in the previous section.

Not all of the Goldstone bosons we have introduced necessarily describe independent degrees of freedom~\cite{Volkov:1973vd,Nielsen:1975hm,Ivanov:1975zq,Low:2001bw}. Depending on the symmetry breaking mechanism, there may be some gauge transformations acting on the Goldstones that do not affect the physical fluctuations of the  order parameter~\cite{Nicolis:2013sga}. When that is the case, one can remove the redundant Goldstone fields by imposing gauge fixing conditions that respect all the 
global symmetries\footnote{These covariant gauge-fixing conditions usually go under the name of ``inverse Higgs constraints". We will use instead the terminology and the interpretation introduced in \cite{Nicolis:2013sga}, which emphasizes their being, in general, optional gauge choices.}.

There is a simple rule of thumb to determine whether such gauge redundancies may exist in the first place \cite{Ivanov:1975zq}. One simply needs to consider the commutators of the unbroken translation generators with a broken symmetry generator. In our case, the relevant commutators~are
\begin{subequations}
\label{commutators}
\be
\left[\p_0, K_i\right]&=&-i(\p_i-Q_i) \, ,\vspace{.1cm}\\
\left[\p_i, F_j\right] &=&-i\delta_{ij}(\p_0-Q_0) \, , \vspace{.1cm} \\
\left[\p_k, M_{ij}\right]&=&-i(\delta_{ik}Q_j-\tfrac{1}{d}\delta_{ij}Q_k)  \, .
\ee
\end{subequations}
Since the broken generators on the RHS's of these equations are independent of the broken generators on the LHS's, the Goldstone fields associated with the latter---namely $\eta^i, \theta^i$ and $\alpha^{ij}$---{\em may} be redundant. Whether they {\em are}, is a question that cannot be answered without further information on the symmetry breaking mechanism \cite{Nicolis:2013sga}. To proceed we will assume that they are, and we will thus construct the effective theory for the minimal set of  Goldstones required to realize all the symmetries. Non-minimal choices where not all the potentially redundant Goldstones are redundant will be studied elsewhere~\cite{framids}.

The gauge fixing conditions that eliminate the redundant Goldstones while preserving all
the symmetries are
\begin{subequations} \la{gf}
\be
&\D_0\pi^i = 0 \, ,& \la{gf1} \\
&\D_i\pi^0 = 0 \, ,&\\
&\D_i \pi_j-\tfrac{1}{d}\delta_{ij}\D_k\pi^k = 0 \, .& \la{gf3}
\ee
\end{subequations}
In $(3+1)$ dimensions, these gauge fixing conditions allow one to eliminate 14 out of 18 Goldstone fields.  \\

\noindent $\boldsymbol{\D_0\pi^i=0}$: We can use the first gauge fixing condition to eliminate the $\eta$ Goldstones in favor of the $\pi$ Goldstones or, equivalently, in favor of the $\phi$'s.    To do so, we use the covariant derivatives given in eqs.~\eqref{Ds}.  It is convenient to parameterize $\Lambda_\mu{}^{\nu}$ in terms of the usual velocity vector $\beta^i$,
and solve  for $\beta^i$ rather for $\eta^i$. Defining the 4-vector $\beta^\mu \equiv (1,-\beta^i)$, we find that $\D_0\pi^i = 0$ implies $\beta^\mu \d_\mu \phi^i =0$. This equation is solved by
\be \la{solbeta}
\beta^\mu = J^\mu/J^0\, ,
\ee
where $J^\mu \equiv \epsilon^{\mu \alpha_1 \ldots \alpha_d}\partial_{\alpha_1}\phi^1\ldots\partial_{\alpha_d}\phi^d$ is the identically conserved current of \cite{Dubovsky:2005xd, Dubovsky:2011sj}. \\

\noindent $\boldsymbol{\D_i\pi^0=0}$: The second gauge fixing condition can be used to eliminate the $\theta$ Goldstones.  Again using equations \eqref{Ds}, we find
\be \la{soltheta}
\theta_i = - \left[ \left( \Lambda \partial \phi \right)^{-1} \right]_i{}^j \Lambda^\nu{}_j \partial_\nu \phi^0.
\ee
where $\Lambda \partial \phi$ stands for $ \Lambda^\mu{}_i \partial_\mu \phi^j$.\\

\noindent $\boldsymbol{\D_i\pi^j-\tfrac{1}{d}\ \delta_i^j\D_k\pi^k= 0}$: The remaining gauge fixing condition can be used to eliminate the $\alpha$ Goldstones.  In this case, it is easier (and sufficient) to solve for $\xi_j^{~i} (\alpha)$. The gauge fixing condition (\ref{gf3}) implies that $\D_k \pi^i \propto \delta_k^i$.  Hence,
\be
\label{X}
\D_k \pi^i =-\delta_k{}^{i}+\Lambda^\nu{}_{k}\partial_\nu\phi^j\,\xi_j{}^{i} = C \delta_k^{i} \, ,
\ee
for some function of the fields $C$.  Solving for $\xi$ gives
\be \la{solxi}
(\xi^{-1})_{k}{}^j = \frac{\Lambda^\nu{}_{k}\partial_\nu\phi^j}{1+C}  \, .
\ee
Finally, using that, by definition, $\det \xi = 1$, we can solve for $C$,
\be\label{X2}
C = -1+\left(\det  \Lambda^\nu_{~k}\partial_\nu\phi^j \right)^{1/d} \, .
\ee
By combining equations (\ref{solbeta}), (\ref{solxi}) and (\ref{X2}), the $\xi_j^{~i} (\alpha)$ can now be expressed solely in terms of the $\phi$'s.\\

Returning to the covariant derivatives \eqref{Ds} for the $\pi$ Goldstones, we can now write them in terms of the $\pi$ (or $\phi$) fields alone.  After imposing the gauge fixing conditions (\ref{gf}), the only non-zero components are $\D_0 \pi^0$ and $\D_1\pi^1=\ldots=\D_d\pi^d =C$. The latter can be simplified using that for any  matrix $A$  one has $\det A= \sqrt{\det A^T A}$, which, combined with the properties of $\beta^i$, yields
\be \la{d1p1}
\!\!\!\!\! \D_1 \pi^1 = (\det \partial_\mu \phi^i \partial^\mu \phi^j)^{1/(2d)} -1 \equiv b^{1/d} -1.
\ee
By combining equations (\ref{dmupi0}), (\ref{solbeta}) and (\ref{soltheta}), we can rewrite also $\D_0 \pi^ 0$ in a fairly compact form:
\be\la{d0p0}
\!\!\! \D_0 \pi^ 0 = \frac{\beta^\mu \partial_\mu \phi^0}{\sqrt{-\beta^\nu\beta_\nu}} -1 = \fr{J^\mu \d_\mu \phi^0}{b} -1 \equiv y -1.
\ee
One may in principle consider also the covariant derivatives of the $\eta$, $\theta$, and $\alpha$ Goldstones, expressed in terms of the $\pi$'s.  However, the solutions (\ref{solbeta}), (\ref{soltheta}) and (\ref{solxi}) show that these fields necessarily start at first order in derivatives of the $\pi$'s, which means that their covariant derivatives can only enter the action at higher orders in the derivative expansion. 

Thus, at lowest order in derivatives, the covariant derivatives (\ref{d1p1}) and (\ref{d0p0}) are the only invariant building blocks of the low-energy action for the perfect fluid, which therefore can be written as
\be
\label{S}
{\cal S} = \int d^D x\, F(b,y) \, ,
\ee
where $F$ is a generic function. Notice that for notational convenience we have removed the fractional $1/d$ power in \eqref{d1p1}, as well as the $-1$ offsets in \eqref{d1p1} and \eqref{d0p0}. This is consistent since $F$ is a completely generic function anyway. We should keep in mind however  that this action should be interpreted as a perturbative series about $b=y=1$, and we have no guarantee that the same effective field theory holds for background values of $b$ and $y$ that are much different than this\footnote{See, however, \cite{Nicolis:2013sga} for certain regularity conditions that can be imposed on the Lagrangian of superfluid systems in order to keep the theory consistent and weakly coupled all the way to zero density.}. 
In many physical systems one will encounter phase transitions at critical values for these quantities. 
Identical considerations apply to the generalizations that we will analyze below.

This action coincides with that  extensively studied in~\cite{Dubovsky:2011sk,Dubovsky:2011sj}. This supports our earlier claim that the fields $\phi^i$ are indeed the comoving coordinates of the fluid. Also, notice that the action \eqref{S} is invariant under the full group of volume preserving diffeomorphisms \eqref{vdiff} as well as the full chemical shift symmetry \eqref{chemshift}, even though only the linearized version of these symmetries was imposed to carry out the coset construction. The reason is that at this order the action only involves the first derivatives of the $\phi$ fields, which---unlike the $\phi$ fields themselves---transform  covariantly under \eqref{vdiff} and \eqref{chemshift}, that is, linearly.


\section{Superfluids}

The coset construction for perfect fluids carried out in the previous section can be easily modified to describe relativistic superfluids. It is well known that a superfluid at finite temperatures can be thought of as an admixture of an ordinary perfect fluid  and a zero temperature superfluid~\cite{landau:1987bo}. From our field-theoretic perspective, this means that a finite-temperature superfluid in $D=d+1$ space-time dimensions can once again be described using $D$ scalars $\phi^\mu (\vec{x},t)$: the $d$ fields $\phi^I$ describe the ordinary component whereas $\phi^0$ describes the superfluid one~\cite{Nicolis:2011cs}. The action must still be invariant under volume preserving diffeomorphisms (\ref{vdiff}), but since the two components do not need to ``flow'' together, the action is only invariant under constant shifts
\be
\phi^0 \rightarrow \phi^0+c^0\, ,
\ee
and not under the full chemical shift (\ref{chemshift}).  The coset construction carried out in the previous section can be easily modified to take this into account.  Setting $\theta_i = 0$ in equation (\ref{Om}) or, equivalently, directly in the covariant derivatives (\ref{Ds}) is formally  equivalent to not having introduced $F_i$ initially.  

The only gauge-fixing conditions we can now impose are those in equations (\ref{gf1}) and (\ref{gf3}). Since both equations involve only the covariant derivatives $\D_\mu \pi^i$, which did not depend on the Goldstones $\theta_i$, their solutions remain the same.  After eliminating the $\beta^i$, the expression for $\D_0 \pi^0$ in terms of the $\pi$ Goldstones also remains unchanged. Thus the Lagrangian for a finite-temperature superfluid is still a function of the quantities $b$ and $y$ defined in equations (\ref{d1p1}) and (\ref{d0p0}). In addition, there is one more invariant that we can write down
\be
\D_i \pi^0 \D^i \pi^0 = \d_\mu \phi^0 \d^\mu \phi^0 -  \frac{(\beta^\mu \partial_\mu \phi^0)^2}{\beta^\nu\beta_\nu} = X + y^2 \, ,~~
\ee
where $X  \equiv \d_\mu \phi^0 \d^\mu \phi^0$.  Therefore, the low-energy effective action for a finite-temperature superfluid is 
\be \la{SsfTnot0}
{\cal S} = \int d^D x\, F(X,b,y) \, \qquad  \mbox{(superfluids, $T\neq0$)} \, , \quad
\ee
in agreement with~\cite{Nicolis:2011cs}.

One can recover the effective action for a superfluid at $T=0$ by neglecting the ordinary component of a finite-temperature superfluid. Following the same logic as above, this amounts to setting $\phi^i =0$ everywhere, in addition to $\theta^i$.  Now there is one covariant derivative of the form
\be
\label{D0}
\D_\mu \pi^0 &=& -\delta_\mu{}^{0}+\Lambda^\nu{}_{\mu}\partial_\nu \phi^0 \, .
\ee
None of our previous commutators \eqref{commutators} can now be used to derive gauge-fixing conditions.  Instead, we note that
\be \label{new commutator}
\left[\p_i, K_j\right] &=&-i\delta_{ij}(\p_0-Q_0) \, .
\ee
Accordingly, we can set $\D_i\pi^0=0$ in order to eliminate the boost Goldstones $\beta^i$ from the spectrum
\footnote{Notice that in the ordinary fluid case we were getting the same $\D_i\pi^0=0$ gauge-fixing condition from the $[\bar P_i, F_j]$ commutator, and we used it to eliminate the $\theta_i$ Goldstones, while the $\beta_i$ Goldstones had been eliminated via the $[\bar P_0, K_i]$ gauge-fixing condition. So, in the ordinary fluid case we did not need the commutator \eqref{new commutator}, because it can be thought of as an {\em alternative} gauge fixing condition for $\beta^i$ with respect to the one we used.}.

We can solve this equation to get: 
\be \la{solbeta2}
\beta_i = \fr{\d_i \phi^0}{\d_0 \phi^0} \; .
\ee
By plugging the relation (\ref{solbeta2}) into the expression for $\D_0 \pi^0$ given above in eq. \eqref{D0}, we get
\be
\D_0 \pi^0 = 1 - \sqrt{-\d_\mu \phi^0 \d^\mu \phi^0} \equiv 1- \sqrt{-X} \, .
\ee
Thus the effective action for the zero-temperature superfluid can be written as
\be
{\cal S} = \int d^D x\, F(X) \, \qquad  \mbox{(superfluids, $T=0$)} \, .
\ee
This agrees with the effective action derived with alternative methods in~\cite{Son:2002zn}.  Notice that the coset construction for superfluids at $T=0$ can be trusted to all orders in the derivative expansion.  This is because the pattern of symmetry breaking involves only a finite number of generators and hence we no longer need to restrict to a smaller subset of generators for the coset construction.

%

%
%


\section{Solids and Supersolids}

The coset construction for perfect fluids can also be adapted to describe relativistic solids and supersolids. Let us start with supersolids. In $D$ space-time dimensions, their low-energy behavior is once again described by $D$ scalar fields $\phi^\mu(\vec{x},t)$~\cite{Son:2005ak}. However, the action for supersolids is invariant only under a subset of the symmetries (\ref{vdiff}) and (\ref{chemshift}), namely (constant) internal shifts and rotations:
\be
\phi^\mu \to \phi^\mu + c^\mu, \qquad \quad \phi^i \to R^i{}_j \phi^j.
\ee
This changes the pattern of symmetry breaking associated with the equilibrium field configuration (\ref{vacuum}). However, after setting $F_i = 0$ and replacing $M_{ij}$ with $L_{ij}$ in equation (\ref{Om}), it is straightforward to repeat the construction carried out in Section \ref{fluids}. The covariant derivatives for the $\pi$ fields are now
\begin{subequations}\label{Ds2}
\be
\D_\mu \pi^0 &=& -\delta_\mu{}^{0}+\Lambda^\nu{}_{\mu} \partial_\nu \phi^0 \, ,\\
\D_\mu \pi^i &=& -\delta_\mu{}^{i}+\Lambda^\nu{}_{\mu}\partial_\nu\phi^j\,R_j{}^{i} \, , \la{Ds2-2}
\ee
\end{subequations}
where $R_j{}^{i} =R_j{}^{i} (\alpha)$ is a $d$-dimensional rotation. The relevant commutation relations to determine potential gauge redundancies are 
\begin{subequations}
\be
\left[\p_0, K_i\right]&=&-i(\p_i-Q_i) \, ,\vspace{.1cm}\\
\left[\p_k, L_{ij}\right]&=&i(\delta_{ik}Q_j-\delta_{jk}Q_i) \,,
\ee
\end{subequations}
and the corresponding gauge fixing conditions are
\begin{subequations}
\be\la{gfSS}
&\D_0\pi^i = 0 \, , \\
&\D_i \pi_j - \D_j \pi_i = 0 \, .
\ee
\end{subequations}
\noindent $\boldsymbol{\D_0\pi^i=0}$: This condition still implies $\beta^\mu \d_\mu \phi^i =0$ and therefore the solution (\ref{solbeta}) remains valid even for supersolids. We can then easily  express the covariant derivatives of $\pi^0$ as a function of the fields $\phi^\mu$ only:
%
\be \la{dss}
\D_0 \pi^ 0 = y -1 \, ,\qquad \qquad 
\D_i \pi^ 0 = \Lambda^\nu{}_i \, \partial_\nu \phi^0 .
\ee
 %

\noindent $\boldsymbol{\D_{[i}\pi_{j]}=0}$: In order to solve this equation to eliminate the $\alpha$ Goldstones, it is convenient to introduce the matrix $N_i{}^j \equiv \Lambda^\mu{}_i \d_\mu \phi^j$. Then, from equation (\ref{Ds2-2}) we see that the condition $\D_{[i}\pi_{j]}=0$ is tantamount to requiring that $N R = S$ with $S$ some symmetric matrix. It follows that $R N = R S R^T \equiv S'$ must also be a symmetric matrix and that $S'^2 = N^TN$. After a few algebraic manipulations, we find that $(S^{\prime 2})_{ij} =\d_\mu \phi_i \d^\mu \phi_j$, which implies:
\be \la{D(IPIJ)}
\D_{(i}\pi_{j)} = - \delta_{ij} + (N\sqrt{ \d_\mu \phi \, \d^\mu \phi} \, N^{-1})_{ij} \, .
\ee
The low-energy effective action for supersolids is therefore a generic functional of the building blocks (\ref{dss}) and (\ref{D(IPIJ)}) that is manifestly invariant under all unbroken symmetries. We note that
\be
\D_i \pi^0 &=&  (N^{-1})^j{}_i \d_\mu \phi^0 \d^\mu \phi_j \, .
\ee
Since all the indices must be contracted to preserve rotational invariance, we can equivalently use as our building blocks 
\be
\begin{array}{lclcl}
\D_i \pi^0& \rightarrow &A_i & \equiv &\d_\mu \phi^0 \d^\mu \phi_i \, ,\\
\D_{(i}\pi_{j)} &\rightarrow& B_{ij}&\equiv &\d_\mu \phi_i \, \d^\mu \phi_j \, .
\end{array}
\ee
Therefore, the low-energy effective action for supersolids is
\be \la{ssaction}
\mathcal{S} = \int d^D x \, F (A_i, B_{jk}, y) \qquad \mbox{(supersolids)}.
\ee
This action is the straightforward relativistic generalization of that derived by Son  in~\cite{Son:2005ak}.

The coset construction for solids is even simpler. The low-energy effective action can be obtained directly by setting $\phi^0 =0$ in the action (\ref{ssaction}). The only remaining building block is then $B_{ij}$, which should be contracted with itself as to preserve rotational invariance. In $D=3$ spatial dimensions, there are only three invariants one can write down using a symmetric matrix such as $B_{ij}$. Following~\cite{Endlich:2012pz}, we can choose them to be
\be
W = [B], \qquad Y = \fr{[B^2]}{[B]^2}, \qquad Z = \fr{[B^3]}{[B]^3},
\ee
where the brackets $[ \cdots]$ stand for the trace of the matrix within. The low-energy effective action for solids is therefore:
\be
\mathcal{S} = \int d^D x \, F (W,Y,Z) \qquad\quad  \mbox{(solids)}.
\ee
%

\section{The $(1+1)D$ Anomaly}
The coset construction performed thus far has the limitation that it only generates terms that are {\it exactly} invariant under the chosen symmetries, as opposed to terms that are invariant up to total derivatives. The latter are known as Wess-Zumino terms and are necessary if one wishes to consider anomalous symmetries. The logic is that, upon gauging, the Wess-Zumino terms may no longer be invariant, thus indicating an anomaly.

There is a straightforward prescription for constructing such terms using the building blocks obtained from the coset construction~\cite{DHoker:1994ti,Goon:2012dy}.  For a $D$-dimensional Wess-Zumino term, one constructs an exact, invariant $(D+1)$-form in $D+1$ dimensions, say $\alpha = d \beta$.  Now, the $D$-form $\beta$ itself is not necessarily invariant but it can shift by a total derivative since $\alpha$ is invariant, and can thus be used as a Wess-Zumino term in the $D$-dimensional action.

As an example, let us consider the Maurer-Cartan form for a perfect fluid in $(1+1)$ dimensions before imposing any gauge fixing conditions.  For now, we will also treat the Goldstone bosons as independent from the space-time coordinates so that we can construct an invariant $3$-form. The Maurer-Cartan form is given by
\be
\Omega^{-1}d\Omega = \omega_{\p}^\mu \p_\mu+\omega_K K+\omega_Q^\mu Q_\mu+\omega_F F \, ,
\ee
where
\begin{subequations} \la{1+1forms}
\be
\omega_{\p}^\mu &=& \Lambda_\nu^{~\mu}\,dx^\nu\, ,  \vspace{.1cm} \\
\omega_Q^0 &=& d\pi^0+\theta \,d\pi^1+(\delta_\nu^{~0}- \Lambda_\nu^{~0}+\theta\delta_\nu^{~1} )dx^\nu \, , \vspace{.1cm} \\
\omega_Q^1 &=& d\pi^1+(\delta_\nu^{~1}-\Lambda_\nu^{~1})dx^\nu \, , \vspace{.1cm} \\
\omega_F &=& d\theta \, ,\vspace{.1cm} \\
\omega_K &=& d\eta \, , 
\ee
\end{subequations}
with $ \Lambda_\nu^{~\mu}\equiv  \Lambda_\nu^{~\mu}(\eta)$. By combining the forms (\ref{1+1forms}) we can write down 20 different 3-forms that are manifestly invariant under the unbroken symmetries, which in $(1+1)$ dimensions are just translations. Out of these 3-forms, we were able to identify 11 linear combinations which are exact. However, only 3 of them give rise to Wess-Zumino terms that have at most one-derivative per field and are exactly Lorentz invariant, rather than invariant up to a total derivative\footnote{We want the Lagrangian to be exactly Lorentz invariant rather than   only up to total derivatives, because the latter option would probably entail gravitational anomalies, i.e.~Lorentz-invariance violations in the presence of gravitational fields, which would be inconsistent with dynamical gravity. Since in the real world gravity {\em is} dynamical, we find it safer to retain exact Lorentz invariance.}.  These are 
\begin{subequations} \la{exact_forms}
\be
&&\omega_{\p}^0 \wedge \omega_{\p}^1 \wedge \omega_Q^1 = d \, ( \pi^1 dx^0 \wedge dx^1 ) \\
&&\omega_{\p}^0 \wedge \omega_{\p}^1 \wedge \omega_F = d \, ( \theta dx^0 \wedge dx^1 ) \\
&&\omega_{\p}^0 \wedge \omega_Q^1 \wedge \omega_F + \omega_Q^0 \wedge \omega_{\p}^1 \wedge \omega_F + \omega_Q^0 \wedge \omega_Q^1 \wedge \omega_F = \nonumber \\
&& = d \, [ \theta ( dx^0\wedge d\pi^1+ d\pi^0\wedge dx^1   +d\pi^0\wedge d\pi^1)].
\ee
\end{subequations}
After pulling back to the space-time manifold and imposing the gauge fixing conditions to express $\theta$ in terms of the $\pi$ Goldstones, we find that the corresponding Wess-Zumino terms are:
\begin{subequations} \la{WZ}
\be
S_{WZ}^{(1)} &=& \int d^2 x \, (\phi^1-1) \\
S_{WZ}^{(2)} &=& \int d^2 x \, \theta = - \int d^2 x \, \fr{\d_\mu \phi^0 \d^\mu \phi^1}{\d_\nu \phi^1 \d^\nu \phi^1} \\
S_{WZ}^{(3)} &=& - \int d^2 x \, \fr{\d_\mu \phi^0 \d^\mu \phi^1}{\d_\nu \phi^1 \d^\nu \phi^1} [\epsilon^{\mu\nu} \d_\mu \phi^0 \d_\nu \phi^1 -1 ] .
\ee
\end{subequations}

The first term is just a tadpole and therefore we will neglect it. The second and third terms are instead more interesting. By design, they are invariant under all the symmetries we assumed as a starting point of our coset construction. However, they are not invariant under the full chemical shift (\ref{chemshift}). From the point of view of our construction, this symmetry arises accidentally at lowest order in the derivative expansion.  Since the Wess-Zumino terms (\ref{WZ}) follow from the forms (\ref{exact_forms}) which have more than one derivative per field, it is not surprising that in general they do not respect this accidental symmetry. What instead is perhaps surprising, is that there still exists a linear combination of $S_{WZ}^{(2)}$ and $S_{WZ}^{(3)}$ that is invariant under the full chemical shift, namely
\be
\label{SWZ}
S_{WZ}^{(2)} + S_{WZ}^{(3)}= -  \int d^2x\,\epsilon^{\mu\nu}\partial_\mu \phi^0 \partial_\nu\phi^1 \frac{\partial_\lambda\phi^1\partial^\lambda \phi^0}{(\partial \phi^1)^2} . \quad 
\ee
This term was extensively discussed in~\cite{Dubovsky:2011sj}.  Our analysis shows that this is the only Wess-Zumino term for a perfect fluid in $(1+1)$ dimensions that is exactly invariant under Lorentz transformations as well as invariant under the full chemical shift, up to a total derivative.

This term \eqref{SWZ} arises from an exact 3-form of the form
\be
\epsilon_{\mu\nu}\, \omega^\mu \wedge \omega^\nu \wedge \omega_F \, ,
\ee
where $ \omega^\mu \equiv \omega_{\p}^\mu+\omega_Q^\mu$.  While it is not straightforward to generalize, this structure may give some hint as to the form of the Wess-Zumino term in higher dimensions, which is still an open question~\cite{Dubovsky:2011sj}.


\section{Outlook}

We have applied the coset construction to the spacetime symmetry breaking patterns characterizing fluid and solid systems.  We thus confirm that the effective field theories that have been considered so far for these systems indeed describe the most general low-energy Goldstone dynamics that are invariant under all the symmetries. 

We plan to apply this formalism to two problems that so far have resisted a satisfactory resolution at the effective field theory level. The first is how to include hydrodynamical dissipative effects systematically. Ref.~\cite{Endlich:2012vt} argues that this should be done by coupling the Goldstones to another sector that ``lives in the fluid."  This sector stands for the microscopic degrees of freedom that are averaged over by the hydrodynamical description, and which are physically responsible for dissipation. The lowest-order couplings have been written down, and they successfully reproduce the standard dissipative effects due to shear viscosity, bulk viscosity, and heat conduction, as well as the associated Kubo relations. It is not obvious, however, how to go beyond linear order in the Goldstones and, more importantly, how to systematically implement the symmetries, given that these ``successful" lowest order couplings {\em violate} some of the symmetries \cite{Endlich:2012vt}. We believe that the coset construction---which provides systematic rules for how to couple the Goldstones to other degrees of freedom in all ways allowed by symmetries and to all orders in perturbation theory---will shed light on these issues. 

The second problem we have in mind is how to incorporate in the Goldstone effective theory Wess-Zumino terms that correctly describe quantum anomalies in (3+1)-dimensional hydrodynamics \cite{Son:2009tf}. Ref.~\cite{Dubovsky:2011sj} constructed a Wess-Zumino term for (1+1)-dimensional fluids carrying an anomalous charge. We reproduced that term above via the coset construction, and showed that in fact it is the only possible Wess-Zumino term consistent with all the symmetries. However, the $(3+1)$-dimensional case is qualitatively more complicated, because it requires moving to one higher order in the derivative expansion, since the anomaly is expected to manifest itself at the one-derivative level beyond the perfect fluid approximation \cite{Son:2009tf}. Notice that this is the same order in the derivative expansion at which dissipative effects appear. It might well be that the two problems are related---in particular, that the insistence that has been placed so far on the infinite-dimensional symmetries \eqref{vdiff} and \eqref{chemshift} is misguided. On the one hand, the linear couplings of \cite{Endlich:2012vt} correctly reproduce the Kubo relations for first order transport coefficients, yet they violate \eqref{vdiff} and \eqref{chemshift}. On the other hand, one can show that the symmetry \eqref{chemshift} implies the existence of a conservation law that seems to be incompatible with anomalous hydrodynamics \cite{son}. It is thus entirely possible that these symmetries have to be abandoned beyond the lowest order in the derivative expansion.   However, one should make sure that there are enough symmetries left that make them re-emerge as accidental symmetries at low enough momenta, for instance as is the case for our ``linearized" symmetries $M_{IJ}$ and $F_I$. 

From this viewpoint superfluids, solids, and supersolids are cleaner systems: they realize finite-dimensional symmetries, and we expect all of these to survive to all orders in the derivative expansion. It would be interesting per se---and useful as a warmup for the fluid case---to use our coset construction to characterize dissipation and anomalies in these systems. We plan to carry out all these projects in the near future.



\vspace{1cm}

\noindent {\bf Acknowledgments:}   This work was supported by NASA under contract NNX10AH14G and by the DOE under contract DE-FG02-11ER41743.

\bibliographystyle{apsrev4-1}
\bibliography{fluids}

\end{document}